 \documentclass[prb,twocolumn,aps,showpacs,amsmath,amssymb,citeautoscript]{revtex4}

\usepackage[dvips]{graphicx}
\usepackage{dcolumn}
\usepackage{bm}

\begin{document}


\title{  
Electron Transport Through Molecules: 
Gate Induced Polarization and Potential Shift
}

\author{ San-Huang Ke,$^{1,2}$ Harold U. Baranger,$^{2}$ and Weitao Yang$^{1}$}

\affiliation{
     $^{\rm 1}$Department of Chemistry, Duke University, Durham, NC 27708-0354 \\
     $^{\rm 2}$Department of Physics, Duke University, Durham, NC 27708-0305
}

\date{\today}

\begin{abstract}
We analyze the effect of a gate on the conductance of molecules by separately evaluating the gate-induced polarization and the potential shift of the molecule relative to the leads.  The calculations use {\it ab initio} density functional theory combined with a Green function method for electron transport.  For a general view, we study several systems: (1) atomic chains of C or Al sandwiched between Al electrodes, (2) a benzene molecule between Au leads, and (3) (9,0) and (5,5) carbon nanotubes.  We find that the polarization effect is small because of screening, while the effect of the potential shift is significant, providing a mechanism for single-molecule transistors.
\end{abstract}

\maketitle


Electron transport through molecules sandwiched between two metallic electrodes has been attracting increasing attention both for fundamental reasons and because it may form the basis of a future molecular electronics technology \cite{mol2,mol3}.  To control the transport and to realize single-molecule-based transistors, a gate is usually applied to the two-terminal lead-molecule-lead (LML) system.  This gate terminal can be (1) global -- i.e. much larger than the molecular region -- made by constructing a plate capacitor near the device \cite{mceuen1,park1,lieber,fuhrer}, or (2) local, grown, for instance, by using chemical vapor deposition beneath the device region \cite{marcus} or simply realized by the sharp tip of a scanning probe microscope \cite{mceuen2}.

The application of a gate to a LML system has two conceptually distinct consequences. First, the voltage on the gate will create a strong electric field in the direction perpendicular to the transport. Second, because of the field component along the transport direction, there will be a potential shift in the molecular region with respect to the leads.  Both effects may influence significantly the electron transport through the molecule.  

Theoretically, a tiny local gate may be modeled in {\it ab initio} calculation by a finite flat surface with a constant electrostatic potential.  Previous {\it ab initio} calculations have evaluated the polarization effect of a local gate for one case \cite{diventra} by puting the molecule into a plate capacitor, and the total effect of a local gate in two others \cite{GuoTaylor01,DattaLiang04} by solving the Poisson equation for the molecular system including the gate.  However, in most recent experiments a golbal gate is used, and its polarization electric field can be quite different from that of a tiny local gate. Additionally, the issue of which effect of the gate is dominant has not been addressed, either experimentally or theoretically. Thus, the individual contributions of the two effects remain unclear.

In this paper, we investigate these two effects -- polarization and potential shift of a global or local gate -- separately from first principles.  The approach is through the well-known combination of density functional theory (DFT) \cite{dft2} for the electronic structure with the non-equilibrium Green function (NEGF) method \cite{negf1,negf2} for electron transport.  Our implementation was developed previously for two terminal systems \cite{trank}.  We use periodic boundary conditions (PBC) for the DFT calculation,  as a result, the geometry of the LML system is accurate without any artificial surface effects.  Using this method, we calculate the molecular conductance under an external polarization electric field or a potential shift in the molecular region, both induced by a gate.

\begin{figure}[b]
\includegraphics[angle=0,width=6.0cm]{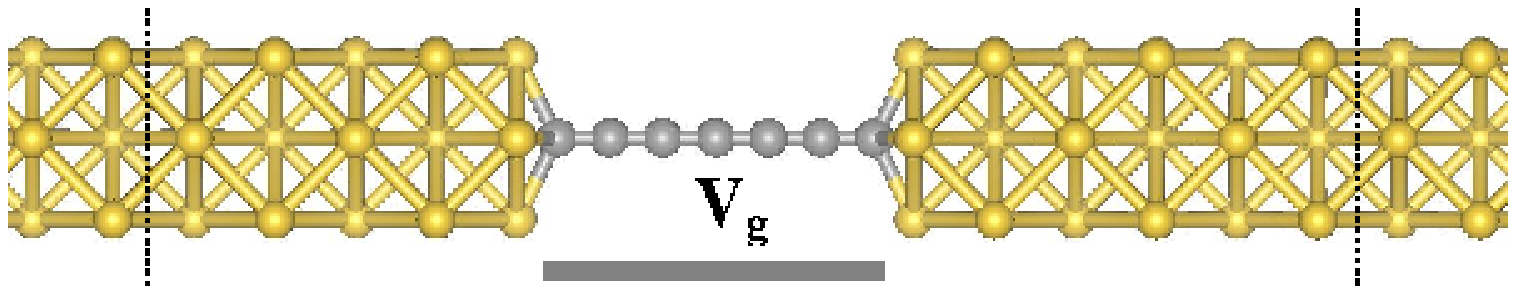} (a)
\includegraphics[angle=0,width=7.6cm]{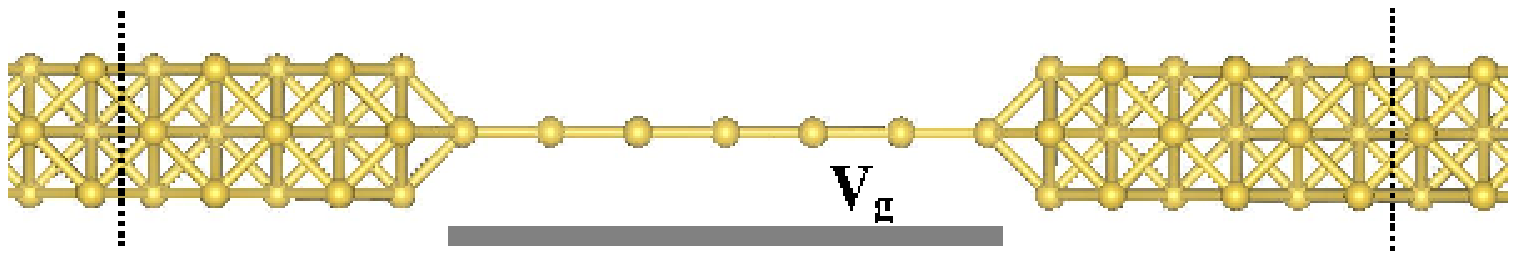} (b)
\includegraphics[angle=0,width=6.7cm]{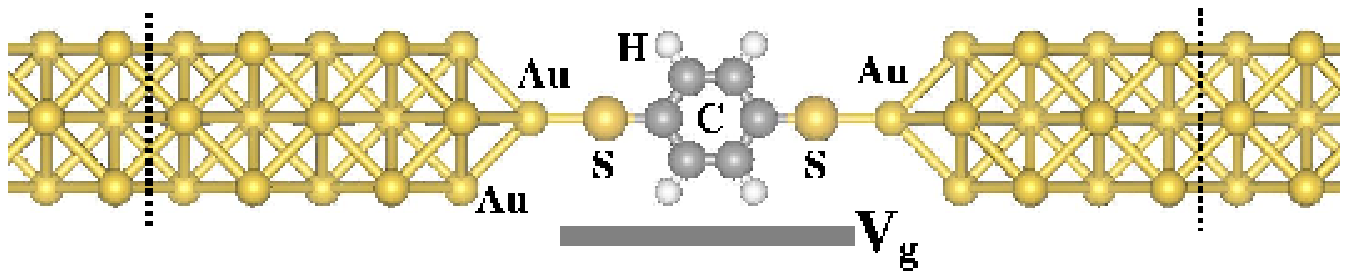} (c)
\includegraphics[angle=0,width=6.2cm]{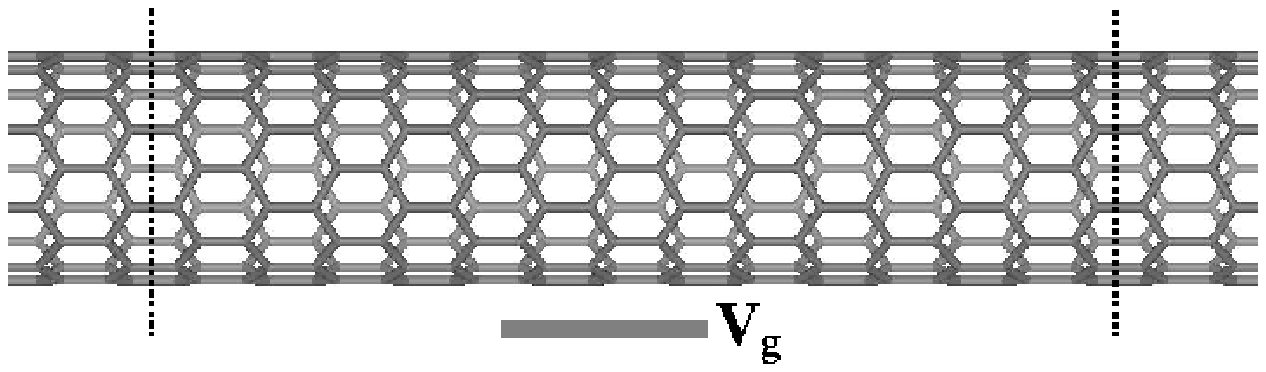} (d)
\includegraphics[angle=0,width=6.2cm]{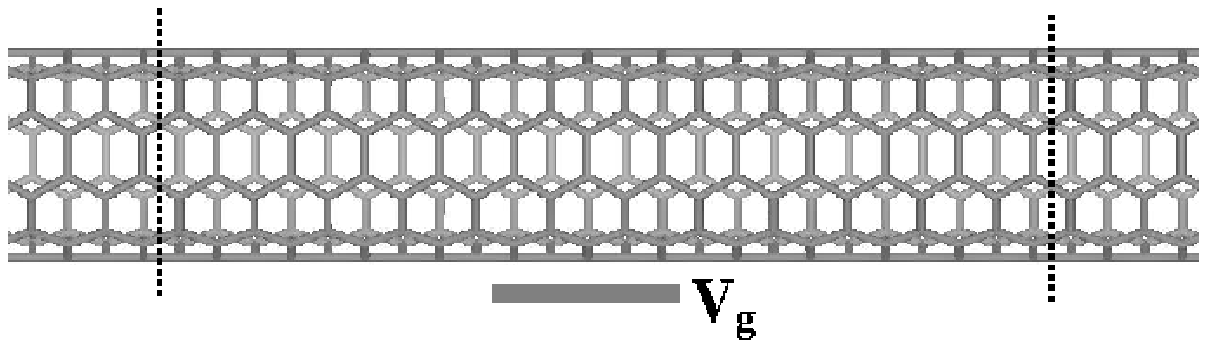} (e)
\caption{Structures of the LML systems calculated:
(a) carbon atomic chain sandwiched between Al(001) leads,
(b) aluminium chain between Al(001) leads,
(c) benzene molecule connected to Au(001) leads through S atoms,
(d) (9,0) carbon nanotube, and (e) (5,5) carbon nanotube.
The dotted line indicates the interface between the device region (extended molecule)
and the leads. The gray bar labeled $V_g$ marks the region where
the potential is shifted by the gate.}
\end{figure}

To arrive at a general view of gate effects, here we investigate three kinds of typical systems: (1) atomic chains, C and Al, sandwiched between two Al(001)--2$\sqrt{2}\times2\sqrt{2}$ leads; (2) a benzene molecule between two Au(001)--2$\sqrt{2}\times2\sqrt{2}$ leads; (3) carbon nanotubes, both semiconducting (9,0) and metallic (5,5).  The atomic structures of these LML systems are shown in Fig. 1.  $V_g$ denotes the local gate potential shift or the locally projected potential shift induced by a global gate.


For zero bias, the density matrix $D$ of the 
device region is simply related to the Green function by
\begin{equation}
\mathbf{D}_{D}=-\frac{1}{\pi}\int_{-\infty}^{+\infty}\textrm{Im} \label{equ_d}
           \left[\mathbf{G}_{D}(E)f(E-\mu_L)\right]dE, 
\end{equation}
where $\mathbf{G}_{D}(E)$ is the retarded Green function of the device region, $f$ is the Fermi function, and $\mu_L$ the chemical potential of the leads.  On the other hand, $\mathbf{G}_{D}(E)$ is determined by the self-energies for the two leads together with the molecular Hamiltonian ($\mathbf{H}_{D}$) given by DFT based on the density implied by $\mathbf{D}_{D}$.  After these two steps converge selfconsistently, the transmission coefficient at any energy, $T(E,V_b)$, is calculated from the Green function. The conductance, $G$, then follows from a Landauer-type relation.

We use a uniform electric field $\mathbf{E}$ to model the field perpendicular to the transport direction caused by a global gate. Because we use periodic boundary conditions for the DFT calculation,\cite{trank} there will be a potential jump at the boundary; this has, however, no unphysical effect since it is in the deep vacuum.  Under the field there is an additional term, $-\mathbf{r} \cdot \mathbf{E}$, in $\mathbf{H}_{D}$.  In the self-energies for the semi-infinite leads, the same uniform electric field is applied.  As a result, the whole infinite LML system will be under a uniform field.  Then, the self-consistency is carried out just as in the zero field case.

To simulate the potential shift along the transport direction induced by a gate, we apply a potential shift directly to the molecular region ($M$),
\begin{equation}
\left[\Delta\mathbf{H}_{D}\right]_{\mu\nu}=V_g \left[
  \mathbf{S}_{D}\right]_{\mu\nu}, \mu \textrm{\;or } \nu \in M, 
\end{equation}
where $\mathbf{S}_{D}$ is the overlap matrix of the device region.  Because the shift is applied to a matrix element when either orbital index is in $M$, the potential shift will be slightly smeared around the molecule-lead contact; we believe this is more realistic than a step-function change.  This potential shift may be applied in a non-selfconsistent (non-SC) or selfconsistent (SC) way.  In the non-SC scheme the shift is added to the initially converged $\mathbf{H}_{D}$ once without further iterations. In the SC scheme $\Delta\mathbf{H}_{D}$ is added to the initially converged $\mathbf{H}_{D}$ for each iteration until $\mathbf{H}_{D}$ and $\mathbf{D}_{D}$ is converged.

A gate potential shift will induce charge accumulation in the molecular region, $\Delta Q_M$, given by $\Delta Q_M = \textrm{Tr}_m [ \mathbf{S}_D \cdot \mathbf{D}_D ]$.  Here Tr$_m$ means trace over only the molecule part, and $\mathbf{D}_D$ is determined by the initially converged $\mathbf{H}_D$ plus $\Delta\mathbf{H}_D$ according to Eq.~(\ref{equ_d}). This charge accumulation in the molecular region will induce opposite charge accumulation on the lead surfaces ($\Delta Q_L$).

The difference between the two schemes is that $\Delta Q_L$ is not taken into account in the non-SC scheme while it is included in the SC scheme by the charge normalization process in each iteration. The SC scheme, therefore, preserves the charge neutrality of the device region.  Note that in the non-SC scheme $\Delta Q_M$ is not compensated at all while in the SC scheme $\Delta Q_M$ is compensated fully by $\Delta Q_L$.  (For the SC scheme, we have to incorporate larger parts of the leads into the device region so that $\Delta Q_L$ can be accommodated.)  The real situation obviously lies between these two extremes: $\Delta Q_M$ is compensated partly by $\Delta Q_L$ and partly by the charge accumulated on the gate.  To reveal the difference between these two limits and to check that our approach is reasonable, we carry out calculations with both schemes.

\begin{figure}[t]
\includegraphics[angle= -90,width=6.5cm]{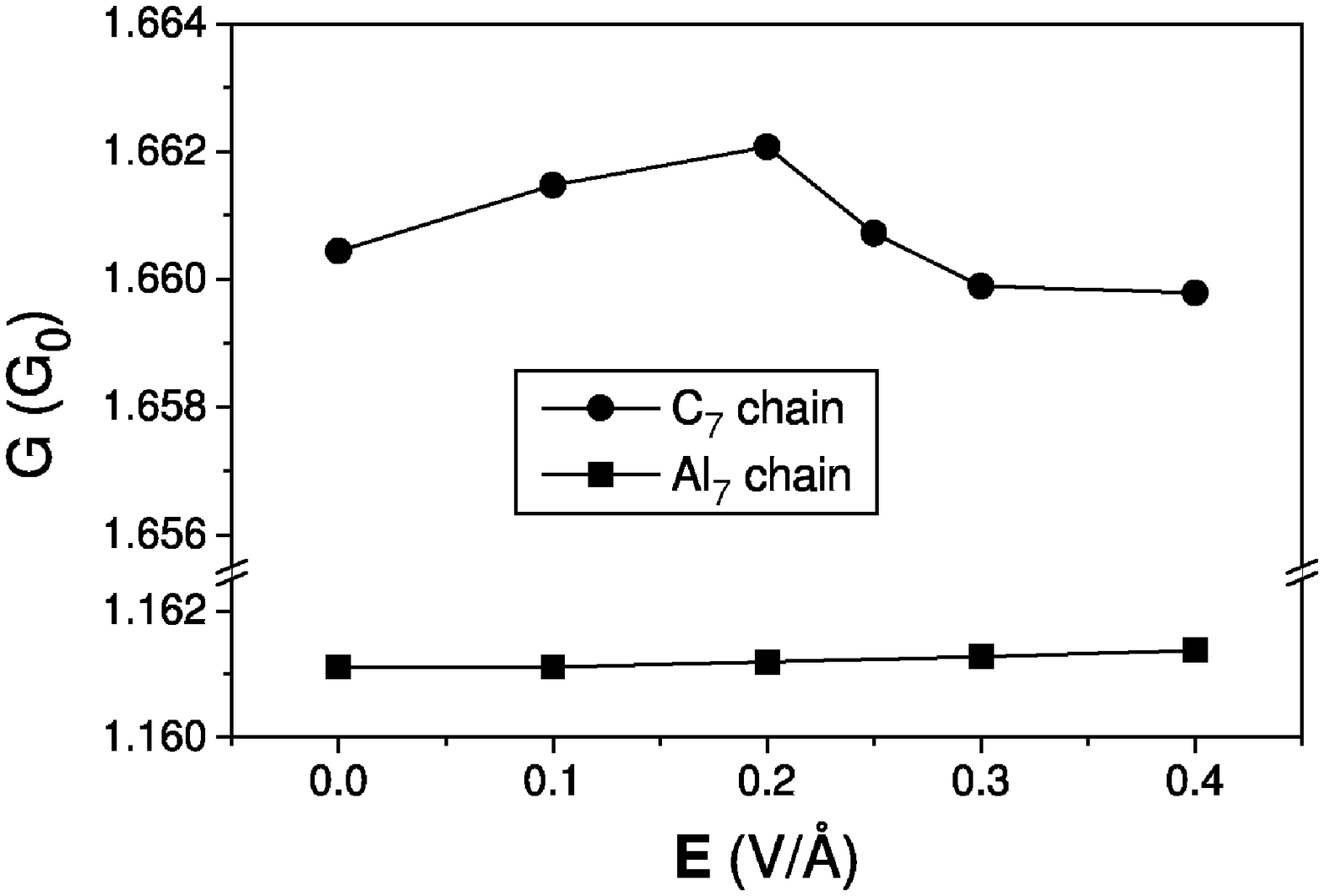} (a)
\includegraphics[angle= -90,width=6.5cm]{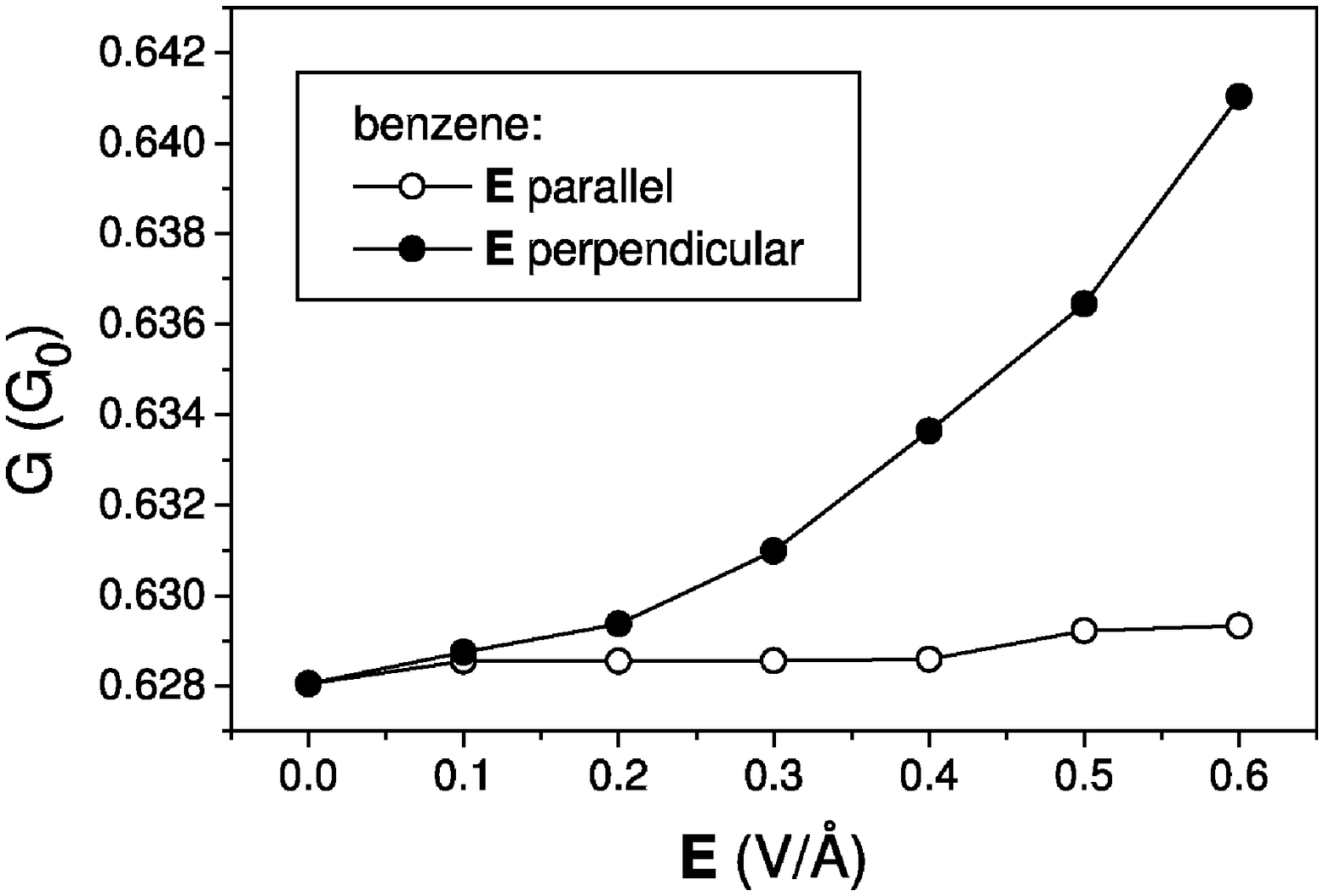} (b)
\caption{Equilibrium conductance as a function of a uniform electric field applied to (a) the C$_7$ and Al$_7$ atomic chains, and (b) the benzene molecule anchored by S atoms.  The direction of the field with regard to the plane of the benzene ring is indicated in (b).  Note the very small variation.}
\end{figure}

\begin{figure}[t]
\includegraphics[angle= -90,width=7.0cm]{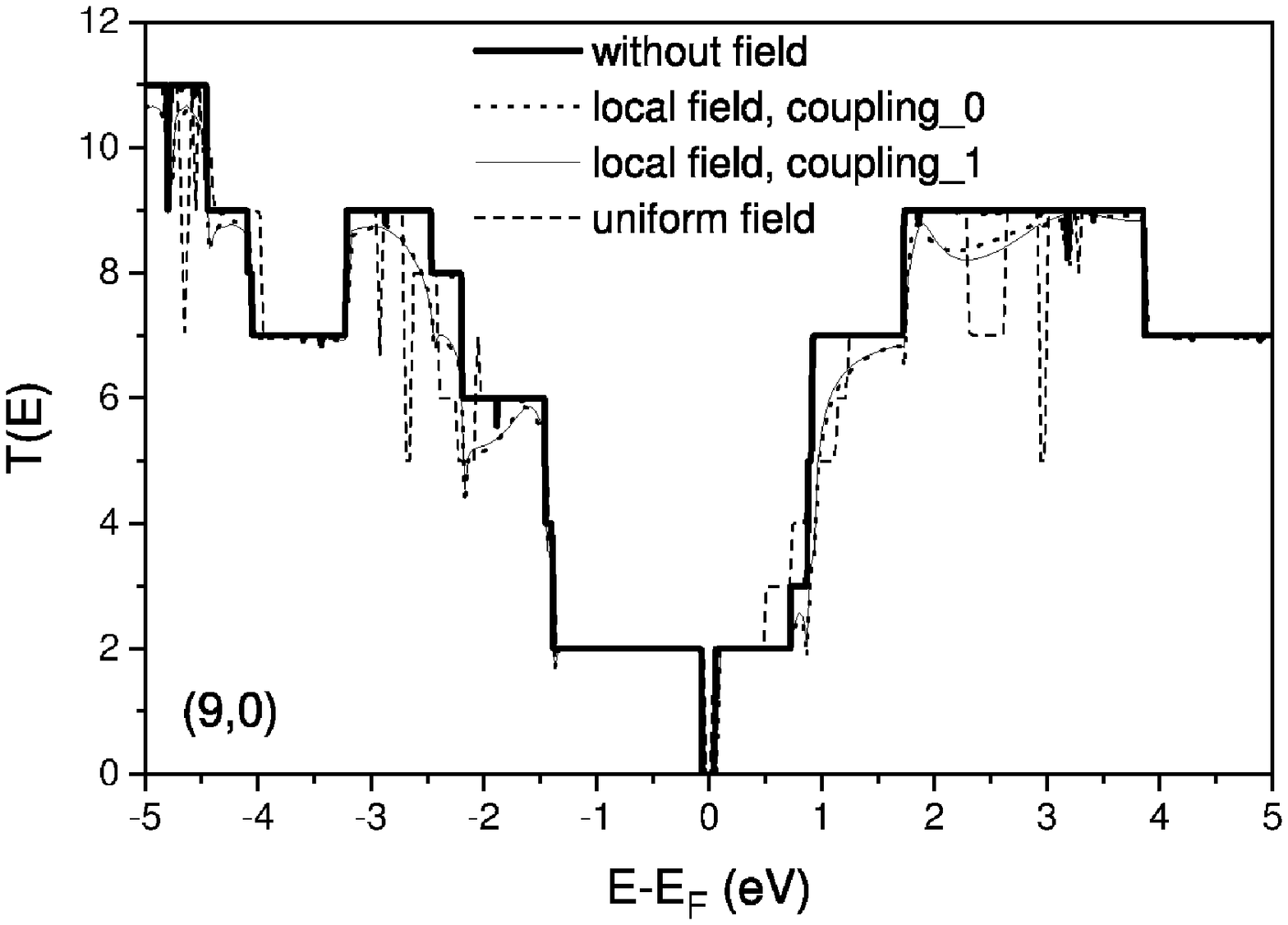} (a)
\includegraphics[angle= -90,width=7.0cm]{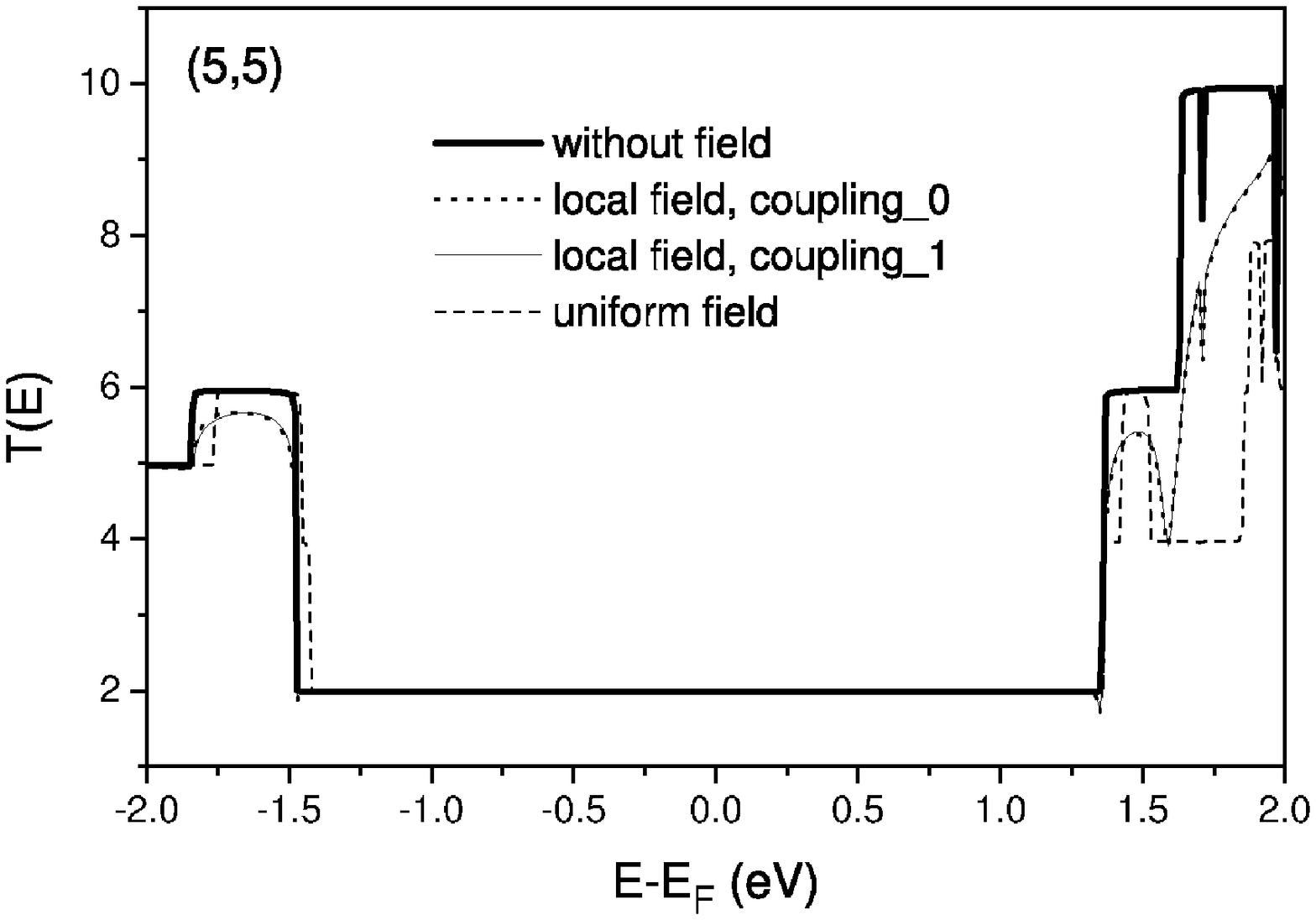} (b)
\caption{Transmission functions of the (9,0) and (5,5) nanotubes
under zero and 0.5 V/{\AA} electric fields. `uniform field' (`local field') means that the whole nanotube (a finite part of
the nanotube) is under the field. `coupling\_0' and `coupling\_1'
indicate the different ways to treat the coupling bwteen the regions
with and without the field, as explained in the text.
Note that the effect is very small around the Fermi energy.}
\end{figure}

\begin{figure*}[t]
\includegraphics[angle= -90,width=7.5cm]{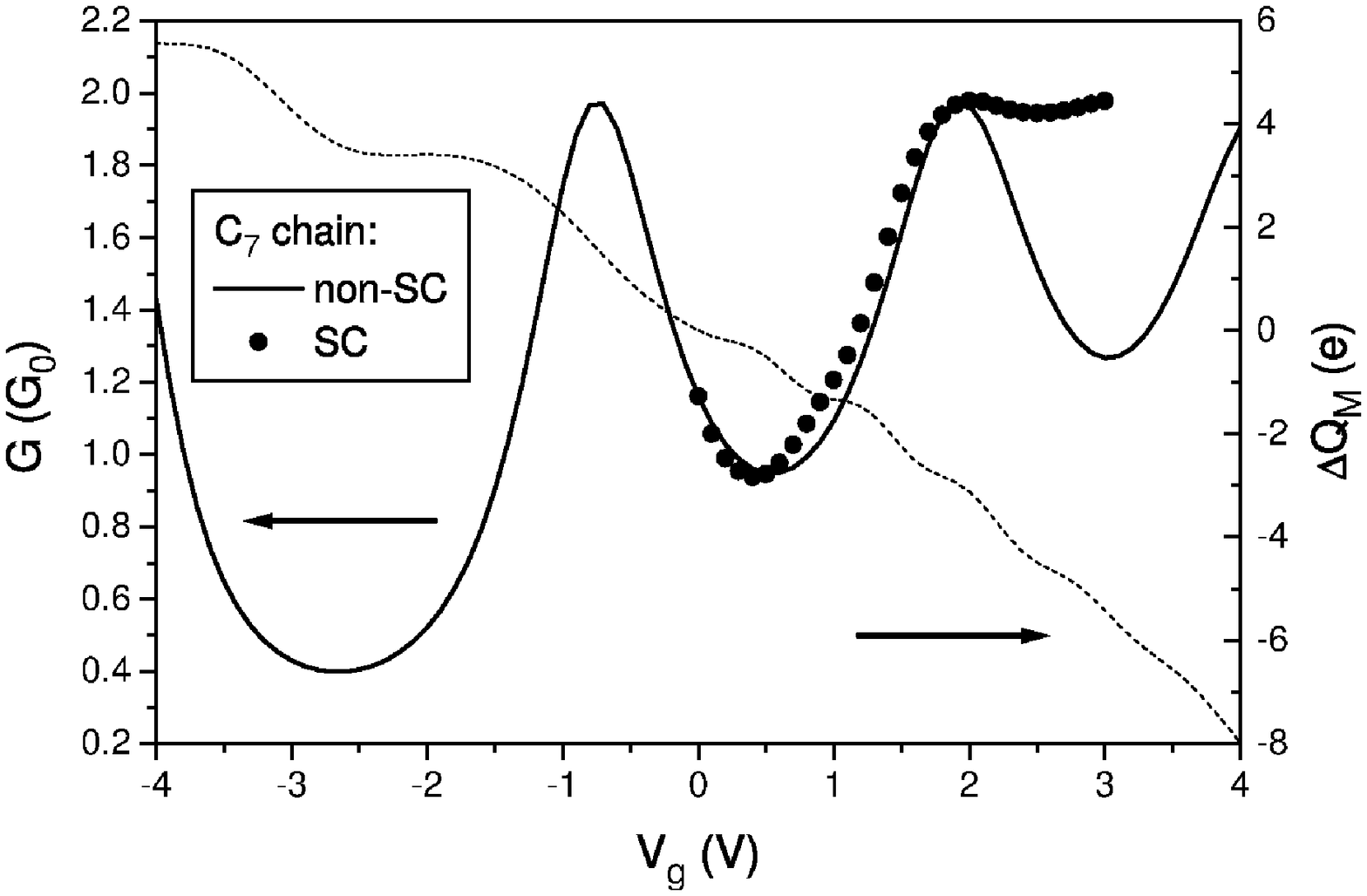} (a)
\includegraphics[angle= -90,width=7.5cm]{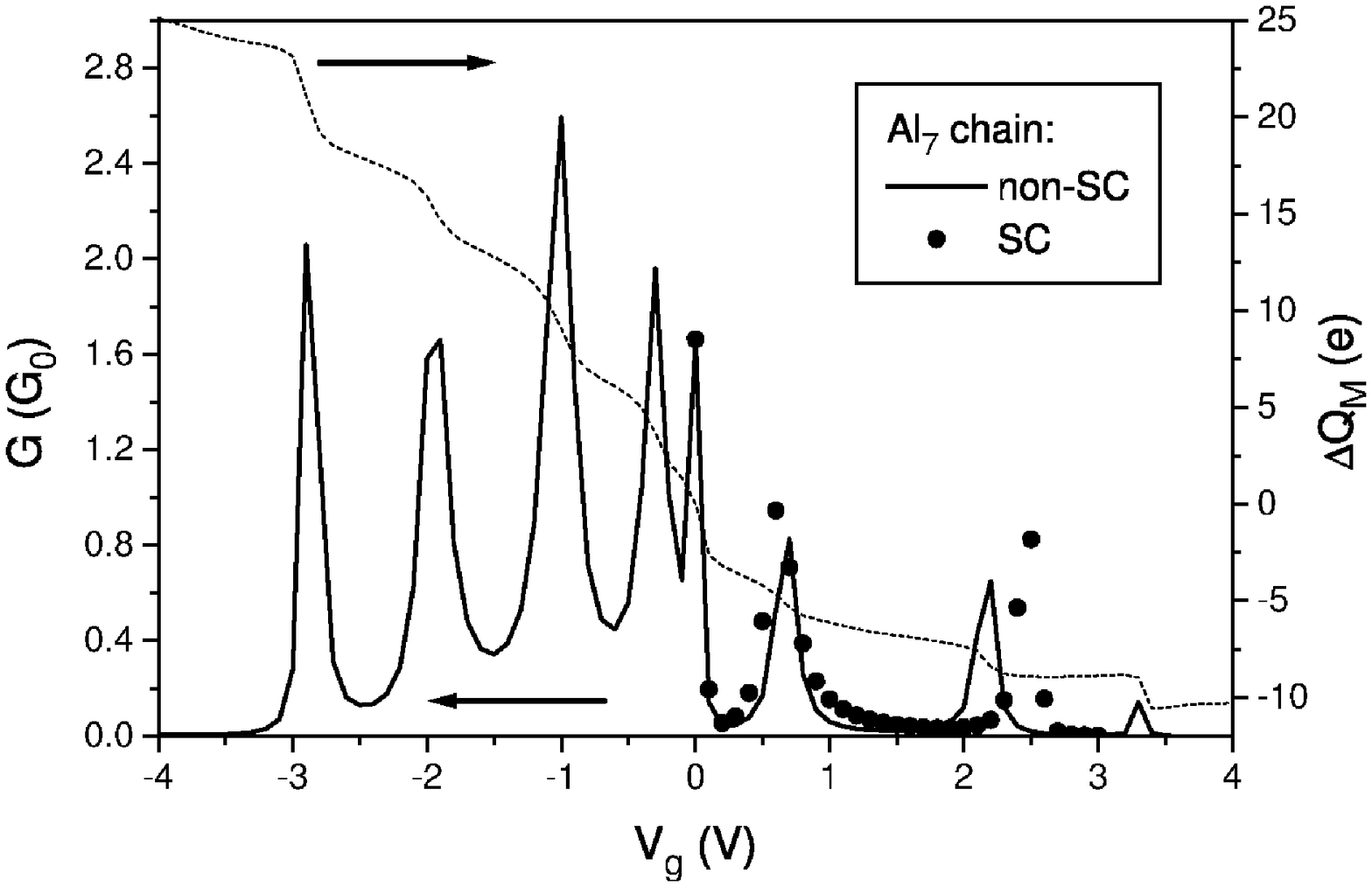} (b)
\includegraphics[angle= -90,width=7.5cm]{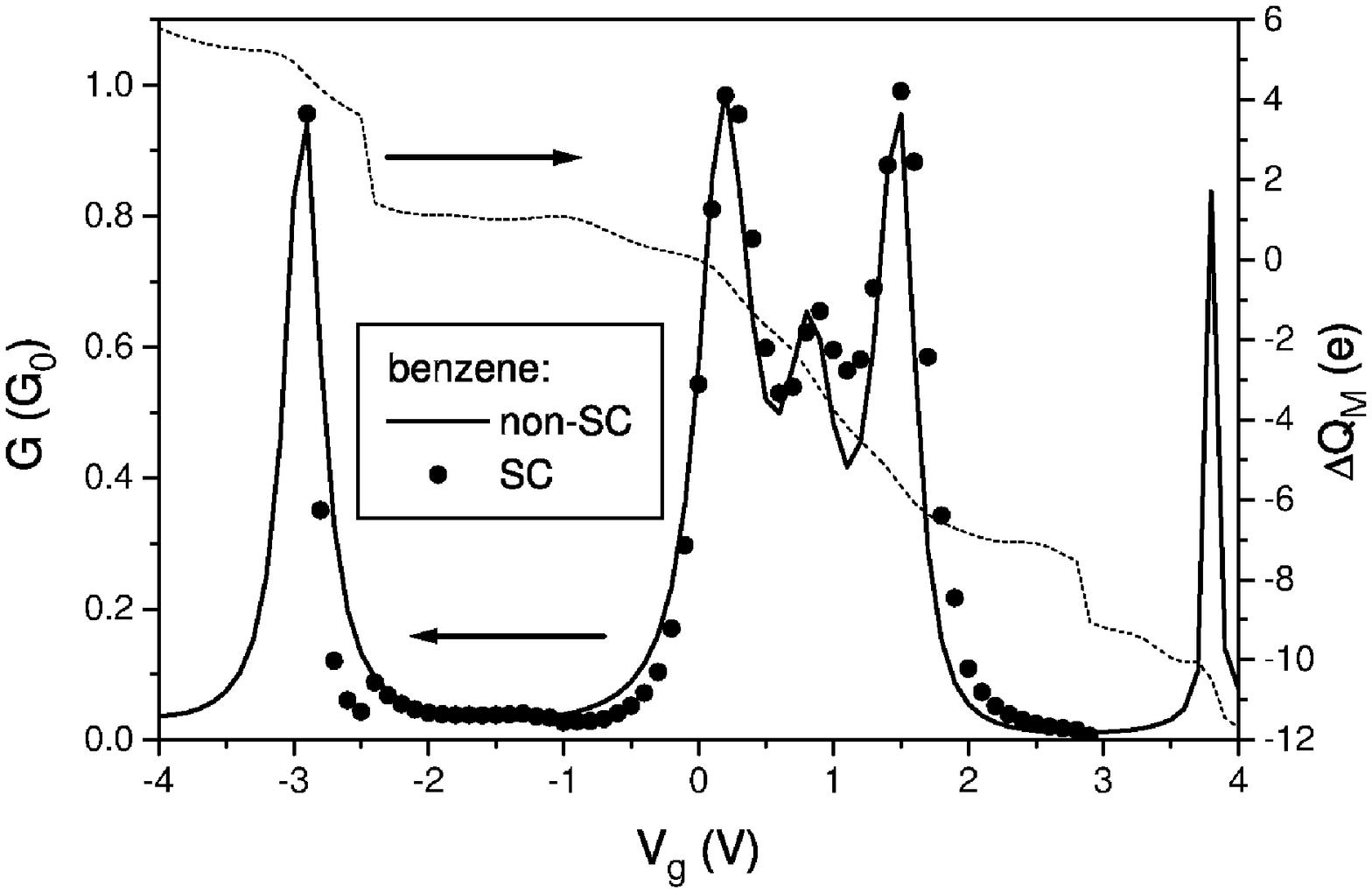} (c)
\includegraphics[angle= -90,width=7.5cm]{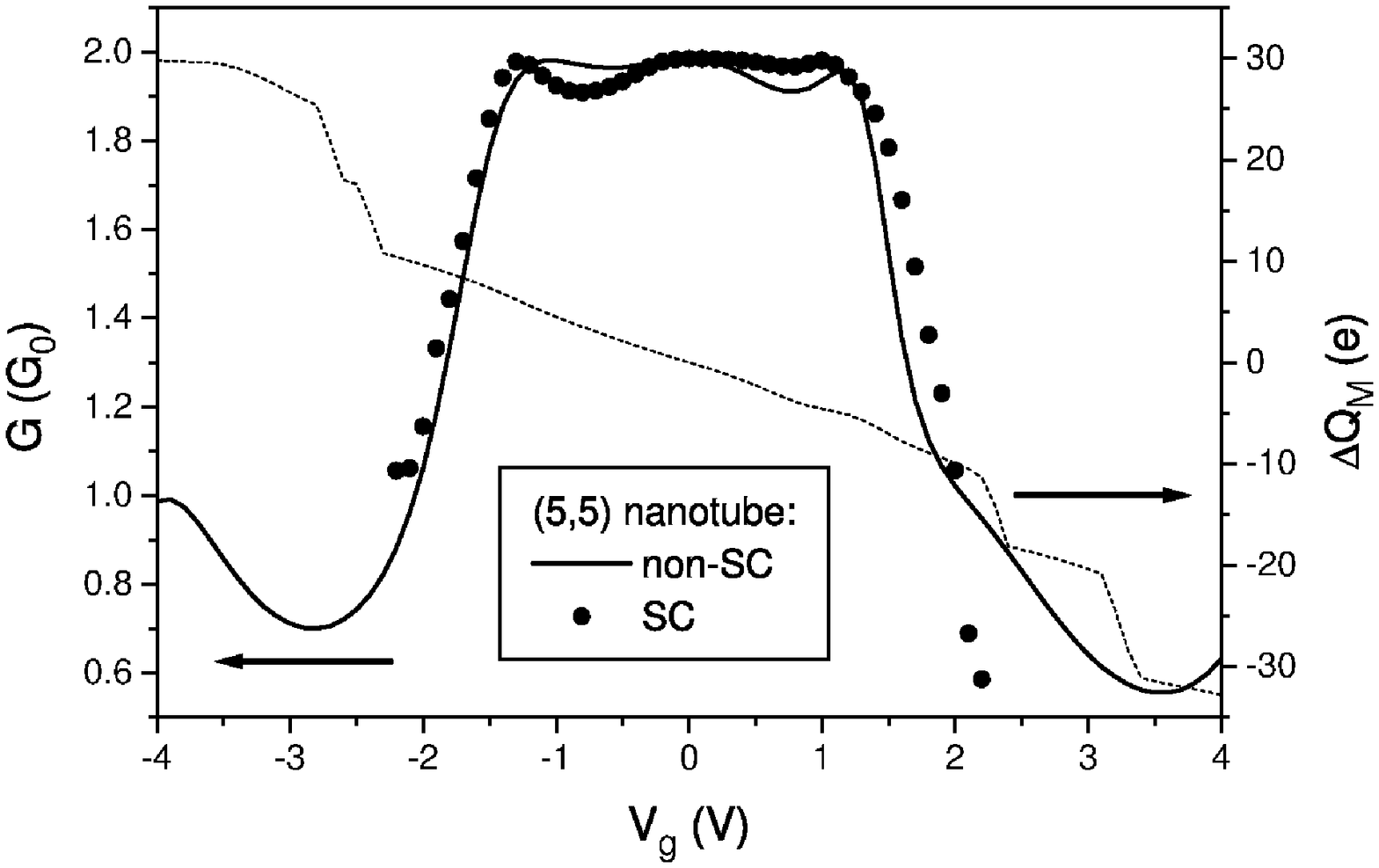} (d)
\caption{Equilibrium conductance as a function of gate potential shift ($V_g$, a positive value means an upward shift of the electron energy) calculated by the non-SC approach (solid line) and SC approach (solid dots), for the four LML systems as indicated.
The charge accumulation in the molecular region ($\Delta Q_M$, in units of electrons) is shown by a dotted line.}
\end{figure*}

We adopt a LCAO-like numerical basis set to expand the wave functions \cite{siesta}, and make use of optimized Troullier-Martins pseudopotentials \cite{tmpp} for the atomic cores.  For the benzene system with gold leads, a high-level double zeta plus polarization basis set is used for all atoms, while for the systems without gold a single zeta basis set is used.  The PBE version of the generalized gradient approximation \cite{pbe} is adopted for the electron exchange and correlation.


In Fig. 2 we show the calculated equilibrium conductance as a function of the global electric field for the atomic chain and benzene systems.  Note that the field applied is very large, approaching the breakdown limit.  Among the atomic chains, the polarization is larger for C than for Al because the screening is stronger in the latter.  For the benzene molecule, the polarization effect is stronger when the field is perpendicular to the ring than when it is parallel, which is related to the characteristics of the large $\pi$ bond.  However, as can be seen, the overall polarization effect is negligible in all cases, indicating that screening plays a very important role in electron transport under a perpendicular electric field.

In Fig. 3 we show the equilibrium transmission functions of the (9,0) and (5,5) carbon nanotubes under zero and 0.5 V/{\AA} electric field.  The field is applied either uniformly (as described above) or locally.  A local field means that it is applied to only part of the device (indicated by the gray bar in Fig. 1) and is not included in the lead self-energies. We treat the coupling between the regions with and without the field in two approximate ways: with or without the field, denoted `coupling\_0' and `coupling\_1', respectively.

For zero field, $T(E)$ shows clear steps because of the quantized electronic states in the directions perpendicular to the transport.  For the metallic (5,5) tube, the equilibrium conductance is 2 conductance quanta ($G_0 =$ $2e^2/h$) -- as is well known, two subbands cross the Fermi energy in this case.  For the (9,0) tube, there is a small gap in $T(E)$ around the Fermi energy -- it becomes semiconducting because of the curvature of the tube.  Under the very strong field of 0.5 V/{\AA}, $T(E)$ still has step structure and is, in fact, almost the same around the Fermi energy.  If this strong field is applied locally, however, then the step structure is destroyed, but $T(E)$ is still almost the same around the Fermi energy.  Thus, the effect of polarization is only felt far from the Fermi energy, indicating that screening plays a very important role just as in the atomic-chain and benzene systems.  

From the similarity of our results for completely different systems we come to a general conclusion: \textit{screening is significant for electron transport through nano-junctions, and, as a result, the effect of a polarizing electric field on transport is small.}

Fig. 4 shows the equilibrium conductance $G$ as a function of gate potential shift for four LML systems, calculated by the SC and non-SC schemes.  Also shown is the charge accumulation in the molecular region, $\Delta Q_M$, due to the gate potential shift.  The first thing to notice is that {\it the equilibrium conductances of all four systems are significantly modified by the gate potential shift}.  For some systems, like the Al chain or benzene molecule, $G$ can vary sharply from near zero to more than one due to only a small change in the gate potential.  This behavior provides a good mechanism for single-molecule transistors.  For benzene, the variation is very sharp just around $V_g=$ $0$; this is related to a resonance near the Fermi energy caused by the two additional Au atoms at the contacts \cite{+Au} [see the contact structure in Fig. 1 (c)].

Different potential shifts lead, of course, to different $\Delta Q_M$. The function $\Delta Q_M(E)$ shows a sort of step-like structure (Fig. 4). In contrast to the Coulomb blockade in which the net charge shows sharp steps, here the steps are smeared out because of the broadening of the molecular orbitals due to the strong molecule-lead coupling.

As mentioned previously, the difference between the non-SC and SC approaches is that $\Delta Q_L$ is not included in the former while probably over included in the latter.  However, it turns out (Fig. 4) that the SC and non-SC results are actually in good agreement for all four systems if the gate potential shift is not too large, $\mid V_g \mid$ $<$ 2 V. Note that for a global gate the potential shift here corresponds to the local projection of the global potential shift, which means that a small $\mid V_g \mid$ can correspond to a much larger global gate potential.  Even for these relatively small potential shifts, $\Delta Q_M$ is already significant, implying that $\Delta Q_L$ should be as well. The good agreement between the SC and non-SC results indicates that the effect of this charge accumulation and the resulting electric field is unimportant for the electron transport.  This conclusion is similar to that for the external electric field above.  This result also justifies that the present SC and non-SC results are reasonable because the real situation lies somewhere between these two extremes.


In summary, we have investigated separately the effects of an external electric field and a potential shift induced by a gate on the molecular conductance of five different lead-molecule-lead systems. We find that the polarization effect of the external field is small because of screening while the potential shift is significant.  The latter provides a good mechanism for single-molecule transistors.

This work was supported in part by the NSF (DMR-0103003).

\end{document}